\documentclass[usegraphicx,useAMS,usenatbib]{mn2e}
\usepackage{times,amsmath,amssymb,afterpage,float}
\usepackage{subfigure}
\title[new periodic methanol maser]{New Periodic 6.7 GHz Class II Methanol Maser Associated with G358.460-0.391}
\author[J. P. Maswanganye et al.]{J.P. ~Maswanganye$^{1,3}$\thanks{Affiliated to HartRAO. E-mail: jabulani@hartrao.ac.za}, M.J. ~Gaylard$^{1}$, S. ~Goedhart$^{2,3}$, D.J. ~van der Walt$^3$ and R.S.~Booth$^{2,4}$ \\
$^1$Hartebeesthoek Radio Astronomy Observatory, PO Box 443, Krugersdorp, 1740, South Africa\\
$^2$SKA SA, 3rd Floor, The Park, Park Rd, Pinelands, 7405, South Africa\\
$^3$School of Physics, North-West University, Potchefstroom campus, Private Bag X6001, Potchefstroom, 2520, South Africa\\
$^4$Department of Physics, University of Pretoria, Private bag X20, Hatfield, 0028, South Africa}

\pagerange{\pageref{firstpage}--\pageref{lastpage}}
\pubyear{2014}
\begin{document}
\date{Accepted 2014 October 29;  Received 2014 October 24; in original form  2014 October 13}

\pagerange{\pageref{firstpage}--\pageref{lastpage}} \pubyear{2014}

\maketitle

\label{firstpage}

\begin{abstract}

Eight new class II methanol masers selected from the 6.7 GHz Methanol Multibeam survey catalogues I and II were monitored at 6.7 GHz with the 26m Hartebeesthoek Radio Astronomy Observatory (HartRAO) radio telescope for three years and seven months, from February 2011 to September 2014. The sources were also observed at 12.2 GHz and two were sufficiently bright to permit monitoring. One of the eight sources, namely G358.460-0.391, was found to show periodic variations at 6.7 GHz. The period was determined and tested for significance using the Lomb-Scargle, epoch-folding and Jurkevich methods, and by fitting a simple analytic function. The best estimate for the period of the 6.7 GHz class II methanol maser line associated with G358.460-0.391 is 220.0 $\pm$ 0.2 day.

\end{abstract}

%

\section{Introduction}

Methanol masers are associated with young, massive star-forming regions, and the brightest class II methanol masers are seen at 6.7 and 12.2 GHz. Some of these masers show aperiodic variability or small to insignificant variations, but, surprisingly, a small proportion show periodic variations which were first reported by \citet{goedhart2003} for G9.621+0.196E. The existence of periodic variations were confirmed in nine more sources by \citet{goedhart2003,goedhart2004,goedhart2007,goedhart2009,goedhart2013}, \citet{araya2010}, \citet{szymczak2011} and \citet{fujiswa2014}. The periodic variations in these masers indirectly reflect the changes in the central massive Young Stellar Object (YSO), or its surroundings, in a manner that is still poorly understood.

Currently, there are four proposals which attempt to explain the origin of the periodic variations in the class II methanol masers. These proposals use the dust temperature \citep{sobolev2007,parfenov2014}, a Colliding Wind Binary (CWB) system \citep{vanderwalt2009,vanderwalt2011}, circumbinary disk accretion \citep{araya2010} or protostellar pulsation \citep{inayoshi2013}. There has been no definitive scientific observation in massive star forming regions with periodic varying class II methanol masers which support any of the above proposals. The CWB proposal, however, seems to be able to describe the waveforms of those sources which show a rapid rise to the maximum followed by an exponential-like decay to the minimum.

The known periodic masers show diverse light curves. Six methanol maser sources show a fairly rapid rise followed by an exponential-like decay to a minimum value. These are G9.621+0.196E, G328.24-0.55, G331.13-0.24, from \citet{goedhart2003,goedhart2004,goedhart2013}, G22.357+0.066 \citep{szymczak2011}, G37.55+0.20 \citep{araya2010} and IRAS 22198+6336 \citep{fujiswa2014}. They can be categorised as having similar flare profiles. From the remaining four periodic sources, G12.89+0.49 and G188.95+0.89 show near sinusoidal type variations, G338.93-0.06 variations were similar to absolute Cosine, $\left| \cos{\left( \omega t\right)} \right|$, and G339.62-0.12 triangular shaped periodic variations \citep{goedhart2003,goedhart2004,goedhart2009,goedhart2013}. 

Given the small number of known periodic masers, the question arises whether there might be more of such cases or these are the only ones. Eight class II methanol masers from the 6.7 GHz Methanol MultiBeam (MMB) survey catalogues I \citep{caswell2010} and II \citep{green2010} were selected for monitoring at 6.7 and 12.2 GHz to partially address the need to find more periodic sources, categorise their light curves and develop a better theoretical understanding of this phenomenon. In the following sections, source selection, observations, data reduction and data analysis techniques are described. The spectra and time series for all masers are shown. One source, G358.460-0.391, showed apparent periodic variations, and this periodicity was tested using four different techniques. Archival data of the methanol maser from the Australia Telescope Compact Array (ATCA) were used to image the maser spots and investigate the maser spot geometry for comparison with known morphological types. Detection rate data from searches for periodic masers were used to estimate the detection rate probability and its uncertainty.
 
%
\section{Observations and data reduction}
%
\subsection{Source selection}

Using the 6.7 GHz Methanol MultiBeam (MMB) survey catalogues I \citep{caswell2010} and II \citep{green2010}, the following steps were used to select the sources for monitoring: (i) The masers were new and (ii) the flux density in either or both epochs was at least greater than ten Jansky, which was to attain at least one hundred peak signal-to-noise-ratio.
 
%
\subsection{HartRAO monitoring}

Eight of the brightest new maser sources from the 6.7 GHz Methanol MultiBeam (MMB) survey catalogues I and II were selected for monitoring and are listed in Table \ref{tab:sources_list}. Note that the sources G358.460-0.393 and G358.460-0.391 are within the primary beam of the telescope, so they were monitored together and can be distinguished by their velocities.

\begin{table*}
 \centering
 \begin{minipage}{140mm}
    \caption{Monitored sources associated with class II methanol masers from the 6.7 GHz MMB survey catalogues I and II. Columns two and three are Right Ascension (RA) and Declination (Dec), respectively, reported by either \citet{caswell2010} or \citet{green2010}. Columns four and five give the velocity range within which masers were found. The flux densities reported from two epochs in the 6.7 GHz MMB survey are also given in column six, from MX mode or beam-switching technique data, and in column seven, from survey cube (SC) data \citep{green2009}.}
  \label{tab:sources_list}
  \begin{tabular}{@{}lccccccc@{}}
  \hline
   Source Name  &  \multicolumn{2}{c}{Equatorial Coordinates}       &   
\multicolumn{2}{c}{Velocity range} & \multicolumn{2}{c}{MMB survey flux} &MMB 
Catalogue Number\\
   ( l,   b )   &  RA (2000)   & Dec. (2000)                        &  $V_L$    
            & $V_H$ (km s$^{-1}$)      & MX data & SC data &                 \\
   ($^{\circ}$, $^{\circ}$) & (h  m  s) & ($^{\circ}$ $^{\prime}$   $^{\prime 
\prime}$) & (km s$^{-1}$) & (km s$^{-1}$) & (Jy)&(Jy) & \\  
 \hline
G0.092-0.663   & 17 48 25.90  &  -29 12 05.9 &  17.0  &  25.0   & 18.86  & 24.80   &  I  \\
G6.189-0.358   & 18 01 02.16  &  -23 47 10.8 & -41.0  &  -26.0  & 228.57  & 221.60 &  II \\
G8.832-0.028   & 18 05 25.67  &  -21 19 25.1 & -12.0  &    7.0  & 159.08  & 126.80 &  II \\
G8.872-0.493   & 18 07 15.34  &  -21 30 53.7 &  21.0  &   29.0  & 33.86   & 27.37  &  II \\
G348.617+1.162 & 17 20 18.65  &  -39 06 50.8 & -25.0  &   -6.0  & 47.59   & 44.26  &   I  \\
G351.688+0.171 & 17 23 34.52  &  -35 49 46.3 & -50.0  &  -30.0  & 41.54   & 42.14  &   I  \\
G358.460-0.391 & 17 43 26.76  &  -30 27 11.3 &  0.0   &    7.0  & 47.73   & 25.40  &   I  \\
G358.460-0.393 & 17 43 27.24  &  −30 27 14.6 & -10.0  &  -6.0   & 11.19   & 14.50  &   I  \\
\hline
\end{tabular}
\end{minipage}
\end{table*}

All sources in Table \ref{tab:sources_list} were observed with the 26m HartRAO telescope at 12.2 GHz in February 2011 to test for the detectability of methanol masers at this frequency. The results showed G8.832-0.028 and G348.617-1.162 had signal-to-noise-ratio of at least one hundred at 12.2 GHz. These sources were observed at least once every three weeks. If a source showed any signs of variability, the sampling cadence was increased to at least once every one to two weeks.

The flux calibrators for the monitoring were Hydra A and 3C123 at both 6.7 and 12.2 GHz; their flux densities were obtained from \citet{ott1994}. Each calibration observation comprised three scans: through the source, and offset at the north and south half power points of the beam. Left- and right-circular polarised outputs were recorded for each scan. The pointing correction was calculated by assuming a Gaussian beam shape. The point source sensitivity (PSS) was derived for left- and right-circular polarisation. The observations were conducted daily, depending on the availability of the telescope.  The average PSS values were used to scale the spectra from antenna temperature (in Kelvin) to flux density (in Jansky).

Spectroscopic observations with the 26m HartRAO radio telescope used a spectrometer providing 1024 channels in each polarisation. The frequency-switching observing technique was used. In the observations, the bandwidth was 1 MHz, providing a velocity resolution of 0.044 km.s$^{-1}$ and typical rms noise in the spectra of 1.2 Jansky. To determine and correct for pointing errors, five scans were made prior to a long on-source observation. Four of these scans were at half power of the beam  in the north, south, east and west directions, and the last one was on-source. The pointing errors were calculated using a Gaussian model for the beam. The weaker sources were not corrected for pointing error owing to the large integration time that would be required for pointing error observations.

After data reduction and calibration, spectra with high pointing errors or high system temperatures, e.g. due to rain, were excluded. The cut-off for the system temperature and pointing error was thirty per cent above the most probable value for both 6.7 and 12.2 GHz. The most probable pointing error was 1.05. In order to reduce the effect of atmospheric absorption and gain variation with hour angle, all observations were made at an hour angle $<$ 3 h.

%
\subsection{ATCA interferometry}

Interferometry data were obtained from the archive of the follow up observations of the 6.7 GHz MMB survey with the ATCA. The project code from the Australian National Telescope Facility (ATNF) archive is C1462. For G358.460-0.391, five snapshot observations were made with six 22m ATCA antennas and each scan was 2.83 minutes long. The primary calibrator PKS B1934-638 was used to do flux scale calibration and the secondary calibrator PKS B1710-269 was used for bandpass and gain calibration. The primary calibrator was observed once at the beginning of the observation for 5.17 minutes. The integration time for the secondary calibrator was 1.67 minutes per scan and ten scans were done. Data reduction was done using Miriad, Common Astronomy Software Applications (CASA) package was used for imaging and Astronomical Image Processing System (AIPS) package was used for determining the position of maser features. 

%
\section{Results}

\begin{figure}
\centering
\resizebox{\hsize}{!}{\includegraphics[clip]{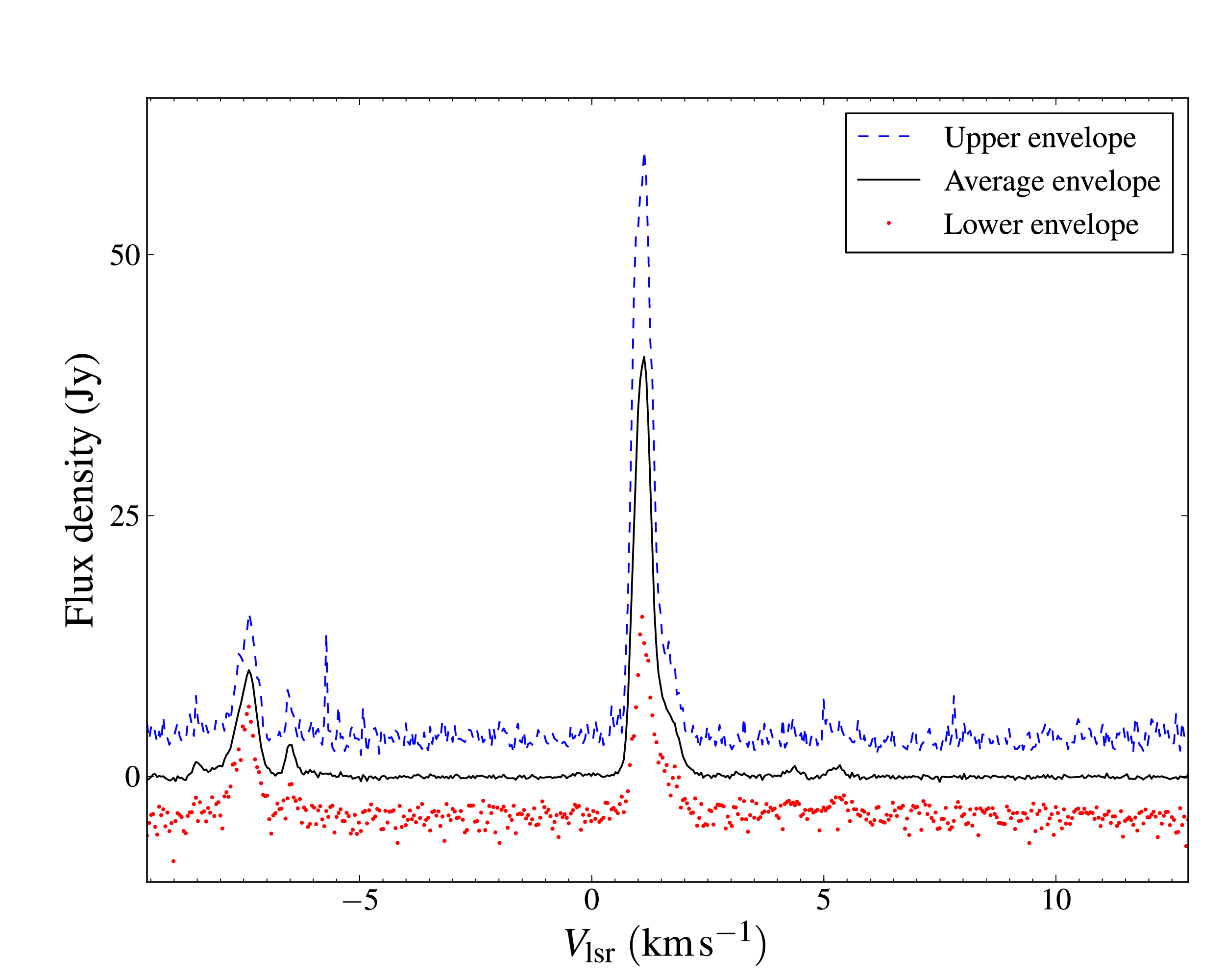}}
\caption{Spectra for the methanol masers associated with G358.460-0.391 and G358.460-0.393 at 6.7 GHz. The left maser feature group is for G358.460-0.393 and
right maser feature is for G358.460-0.391.}
\label{fig:G358.460-0.391_67_spectra}
\end{figure}

The spectra of the class II methanol masers associated with G358.460-0.391 and G358.460-0.393 are shown in Figure \ref{fig:G358.460-0.391_67_spectra}. The spectra are similar to those reported by \citet{caswell2010}. The upper envelope spectra were created by finding the highest flux density attained in each velocity channel and the lower envelopes were the lowest flux density in each channel, over our monitoring period.

\begin{figure}
\resizebox{\hsize}{!}{\includegraphics[clip]{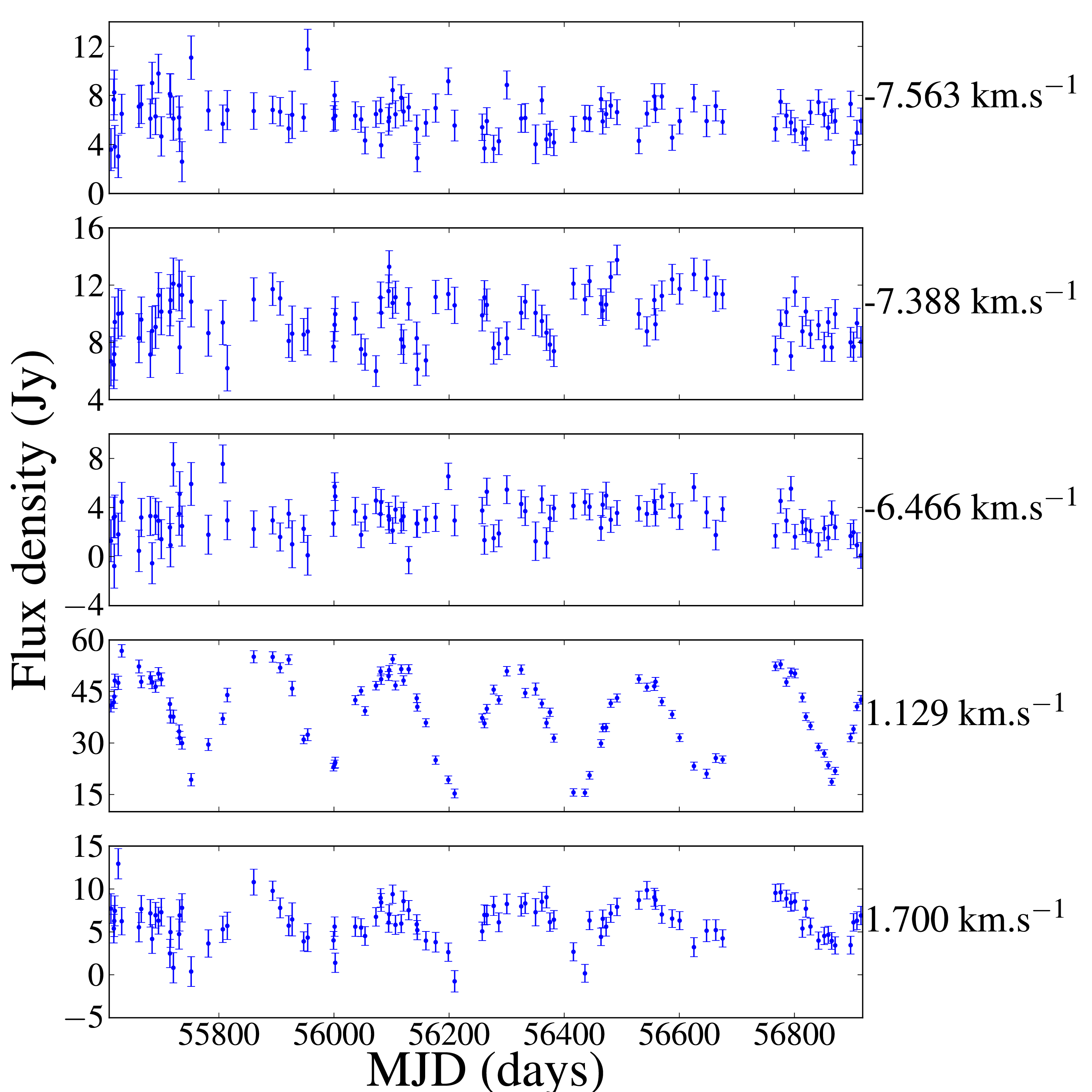}}
\caption{Time series for the methanol masers associated with G358.460-0.391 and G358.460-0.393 at 6.7 GHz. The positive velocities are masers from G358.460-0.391 and the negative velocities are from G358.460-0.393.}
\label{fig:G358.460-0.391_67_timeseries}
\end{figure}

\begin{figure}
\resizebox{\hsize}{!}{\includegraphics[clip]{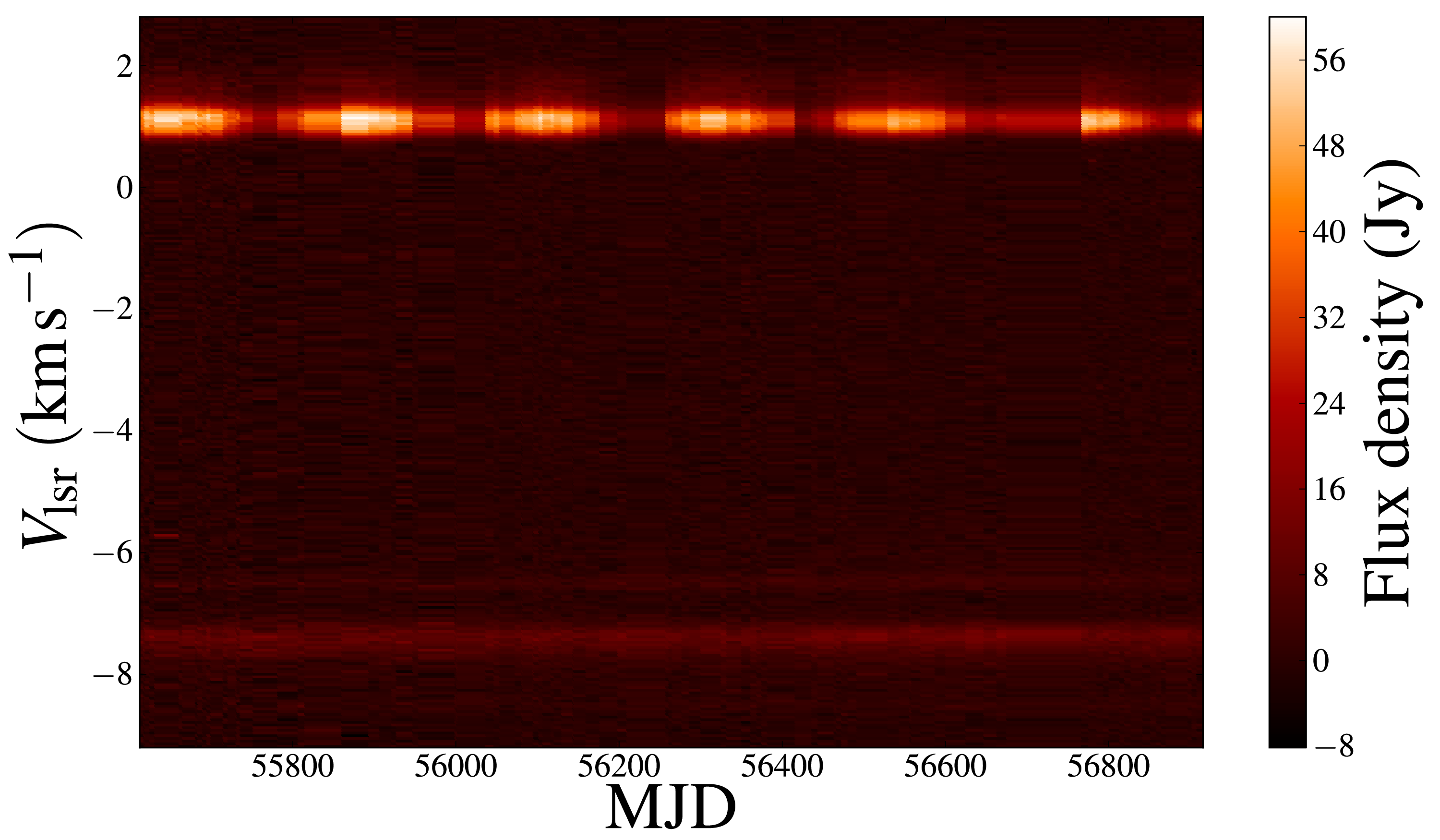}}

\caption{Time series for class II methanol masers associated with G358.460-0.391 and G358.460-0.393 at 6.7 GHz in 2-d contour form. The $-10.0$ to  $-6.0$ km.s$^{-1}$  velocity range is for masers associated G358.460-0.393 and G358.460-0.391 masers are found at $0.0$ to $2.0$ km.s$^{-1}$ velocity range.}
\label{fig:G358.460-0.391_67_2d_timeseries}
\end{figure}

The time series of the selected maser velocity channels associated with G358.460-0.391 and G358.460-0.393 are shown in Figure \ref{fig:G358.460-0.391_67_timeseries}. The maser features were selected on the basis that they reflect the behaviour of the adjacent velocities and they are the brightest. The time series for G358.460-0.391, which produces the maser features at positive velocities, show periodic variations over the monitoring period. The features at negative velocities are from G358.460-0.393 and show no significant variability. Figure \ref{fig:G358.460-0.391_67_2d_timeseries} shows the 2D contour plot for the time series of G358.460-0.391 and G358.460-0.393, and highlights the difference in their behaviour.
     
The spectra and time series for the remaining non-varying six methanol sources in the monitoring programme are shown in the appendix section.


%
\section{Data analysis techniques}

All the sources in our sample were tested for periodicity and its significance using three independent methods: Lomb-Scargle \citep{lomb1976,scargle1982,press1989}, Jurkevich \citep{jurkevich1971} and epoch-folding using Linear-Statistics (or L-statistics) \citep{davies1990}.

The Lomb-Scargle period search method modelled the time series as the sum of a Cosine and Sine. The period is found in the periodogram as the location of a fundamental peak. The significance of the determined period is tested by the false alarm probability method \citep{scargle1982}. 

The epoch-folding period search method produced the phases ($\phi_i$) from the modulus of the trial period ($P$) and they are defined as $\phi_i = \frac{t_i}{P} - [\frac{t_i}{P} ] $, where the square brackets mean only the integer of the number in square brackets is considered. The phases and their corresponding flux densities are binned into $k$ bins. The period can be determined either by the Phase Dispersion Minimization test statistic \citep{stellingwerf1978} or the epoch-folding test statistic \citep{leahy1983}, which was used here, and the significance of the period was tested by L-statistics  \citep{davies1990}.

In the Jurkevich statistics, the time series is folded in the same way as in epoch-folding method for all trial periods. The variance is then calculated from the folded time series, and this reaches its absolute minimum when the time series is folded by the true period. \citet{kidger1992} used the normalised absolute minimum to describe the strength of periodicity in the time series.

%
\section{Analysis}
%
\subsection{Period search}

From the  sample of eight sources, only methanol masers associated with G358.460-0.391 show evidence of periodic variations over the monitoring period. \citet{caswell2010} noted that the flux density observed at two epochs was different, but the source was not reported as a periodic variable.

\begin{figure}
\centering
\resizebox{\hsize}{!}{\includegraphics[clip]{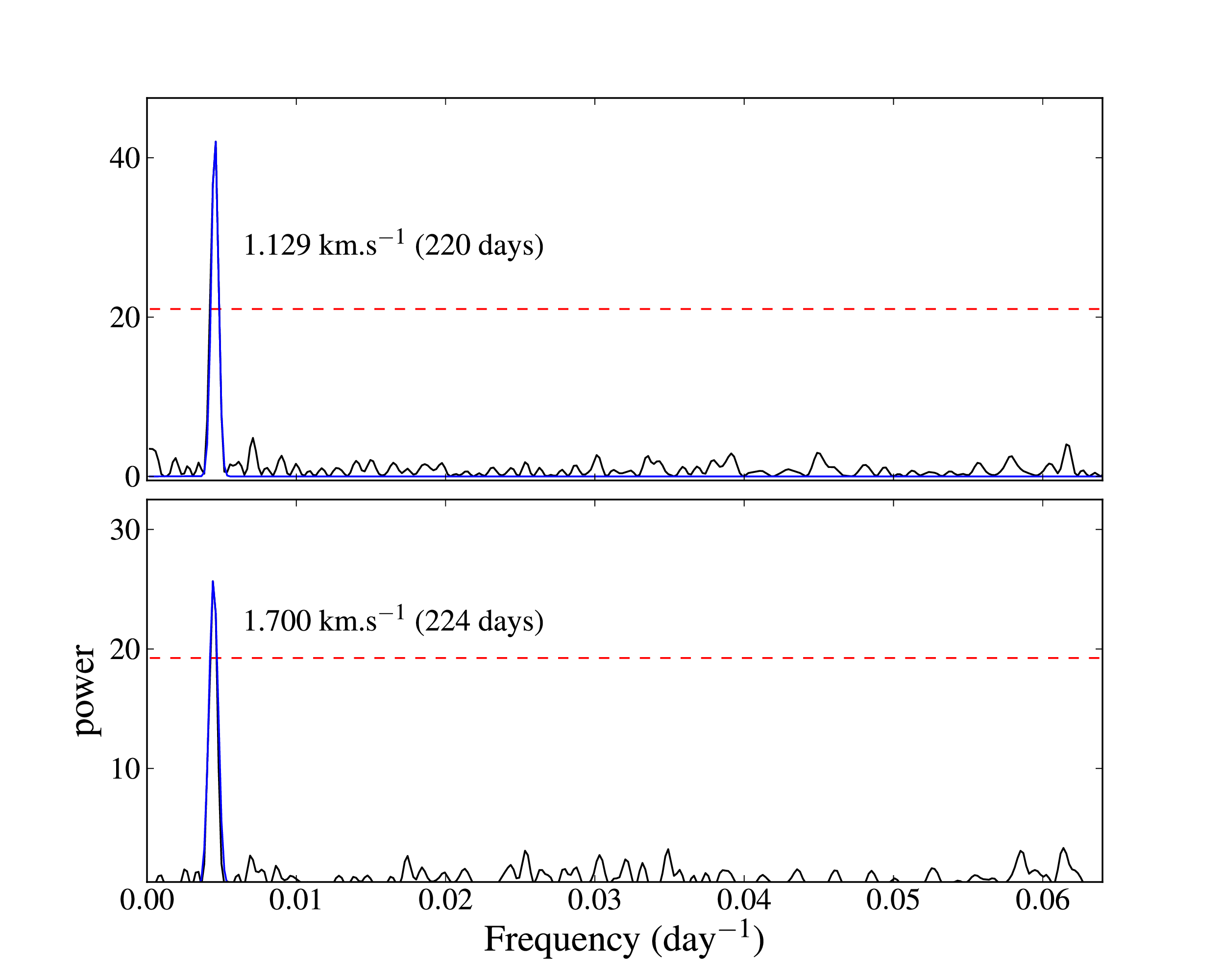}}
\caption{The Lomb-Scargle periodogram for G358.460-0.391 at 6.7 GHz. The dashed line is the cut-off flux for false detected peak. So peaks below the cut-off power are considered to be false alarms or false detection of the peaks.}
\label{fig:G358.460-0.391_lomb_scargle}
\end{figure}

\begin{figure}
\centering
\resizebox{\hsize}{!}{\includegraphics[clip]{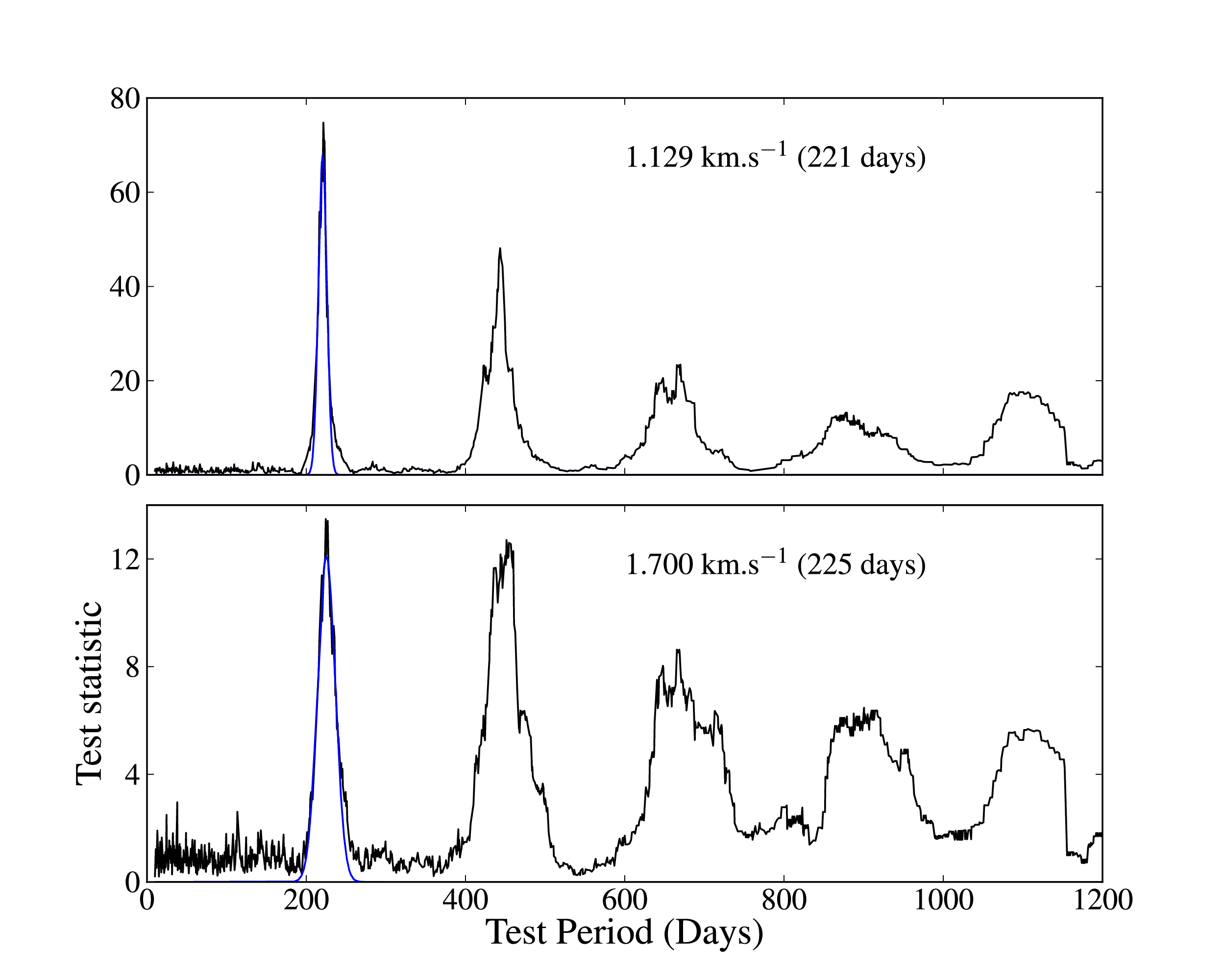}}
\caption{The epoch-folding result using L-statistics for G358.460-0.391 at 6.7 GHz.}
\label{fig:G358.460-0.391_epochfolding}
\end{figure}

\begin{figure}
\centering
\resizebox{\hsize}{!}{\includegraphics[clip]{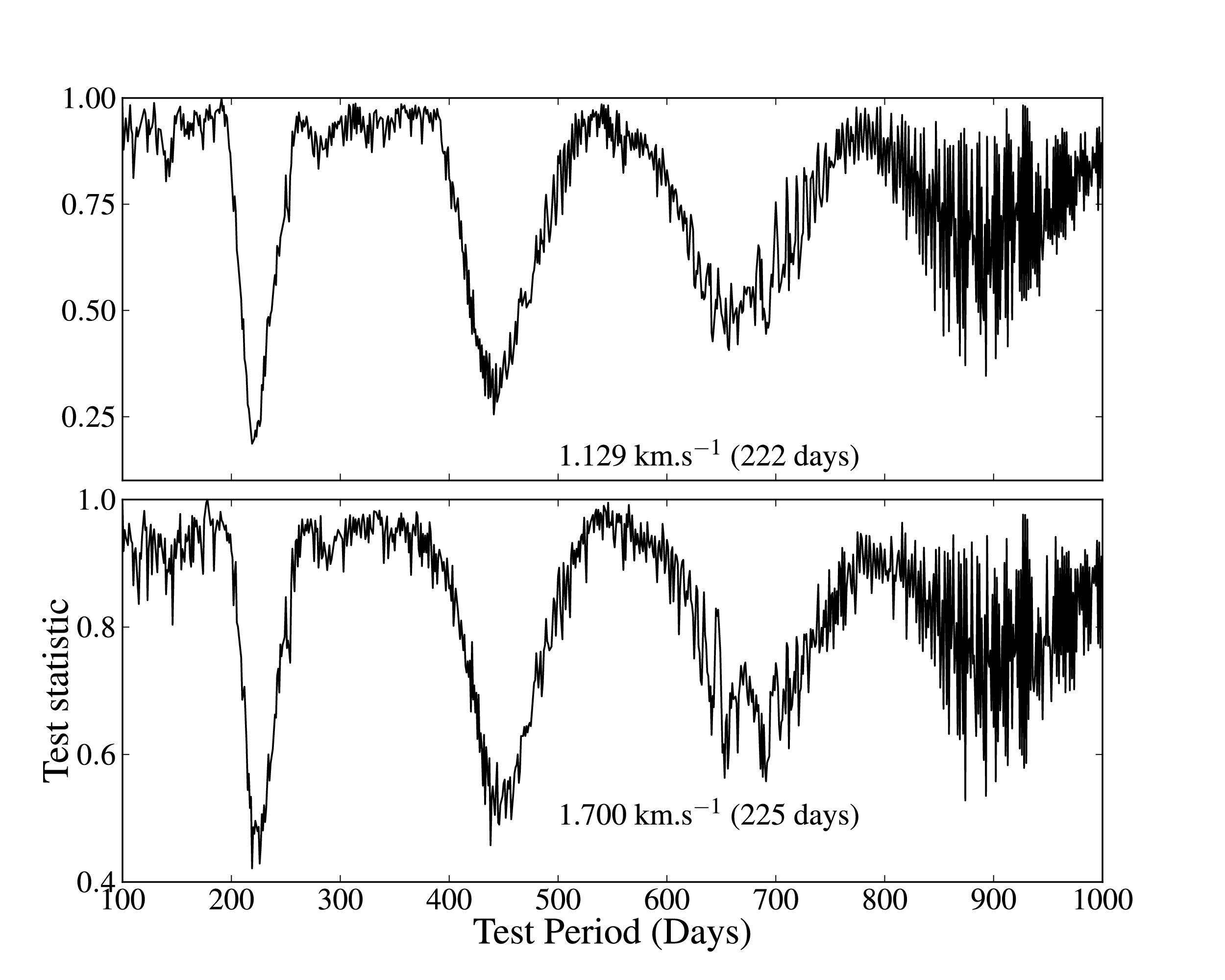}}
\caption{The Jurkevich-statistics result for G358.460-0.391 at 6.7 GHz.}
\label{fig:G358.460-0.391_jurkevich}
\end{figure}

The results of testing the maser time series of G358.460-0.391 (Figure \ref{fig:G358.460-0.391_67_timeseries}) for periodicity are shown for the Lomb-Scargle technique in Figure \ref{fig:G358.460-0.391_lomb_scargle}, epoch-folding method in Figure \ref{fig:G358.460-0.391_epochfolding} and Jurkevich technique in Figure \ref{fig:G358.460-0.391_jurkevich}. The summary of the determined periods is given  in Table \ref{tab:periods}. These three methods independently confirm that G358.460-0.391 shows periodic variations over the monitoring period. The uncertainties in the determined periods were estimated as follows: a Gaussian function was fitted to the top half of the fundamental peak. The period was estimated from the maximum of the fitted Gaussian and the half width at half maximum of the Gaussian (HWHM) provided the error estimate for the period. The Lomb-Scargle, epoch-folding and Jurkevich methods agree in the period and their periodograms have  good signal-to-noise ratios.

\begin{table}
  \caption{Summary of the determined periods using the Lomb-Scargle, epoch-folding and Jurkevich period search methods.}
  \label{tab:periods}
  \begin{center}
  \begin{tabular}{@{}cccc@{}}
  \hline
  Maser feature velocity  & Lomb-Scargle & Epoch-folding & Jurkevich  \\
    (km.s$^{-1}$)   & (days)       &    (days)    &   (days)   \\
    \hline
    1.129           &  220 $\pm$ 14  & 221 $\pm$ 6    & 222 $\pm$ 18 \\
    1.700           &  224 $\pm$ 18  & 225 $\pm$ 13   & 224 $\pm$ 21 \\
    \hline
\end{tabular}
\end{center}
\end{table}

The strength of the periodicity can be calculated using the relation defined by \citet{xei2008}:
\begin{equation}
f = \frac{1 - {V_{m}}^2}{{V_{m}}^2},
\label{eq:periodicity_measure}
\end{equation}
where ${V_{m}}^2$ is the normalised absolute minimum variance in the Jurkevich periodogram. There are three possible outcomes for $f$ which were described by \citet{xei2008}, namely i) $ f = 0$ implies that there is no periodicity and it occurs when ${V_{m}}^2 = 1$; ii) $f < 0.25$ implies weak periodic variation and iii) $f \geq 0.5$ implies the time series is strongly periodic. For the 1.129 km.s$^{-1}$ time series, ${V_{m}}^2 = 0.17$ which translates into $f = 4.9$ and for the 1.700 km.s$^{-1}$,  ${V_{m}}^2 = 0.41$ and $f = 1.4$. This implies that the time series of the masers at both 1.129 and 1.700 km.s$^{-1}$ are strongly periodic.

\begin{figure}
\centering
\resizebox{\hsize}{!}{\includegraphics[clip]{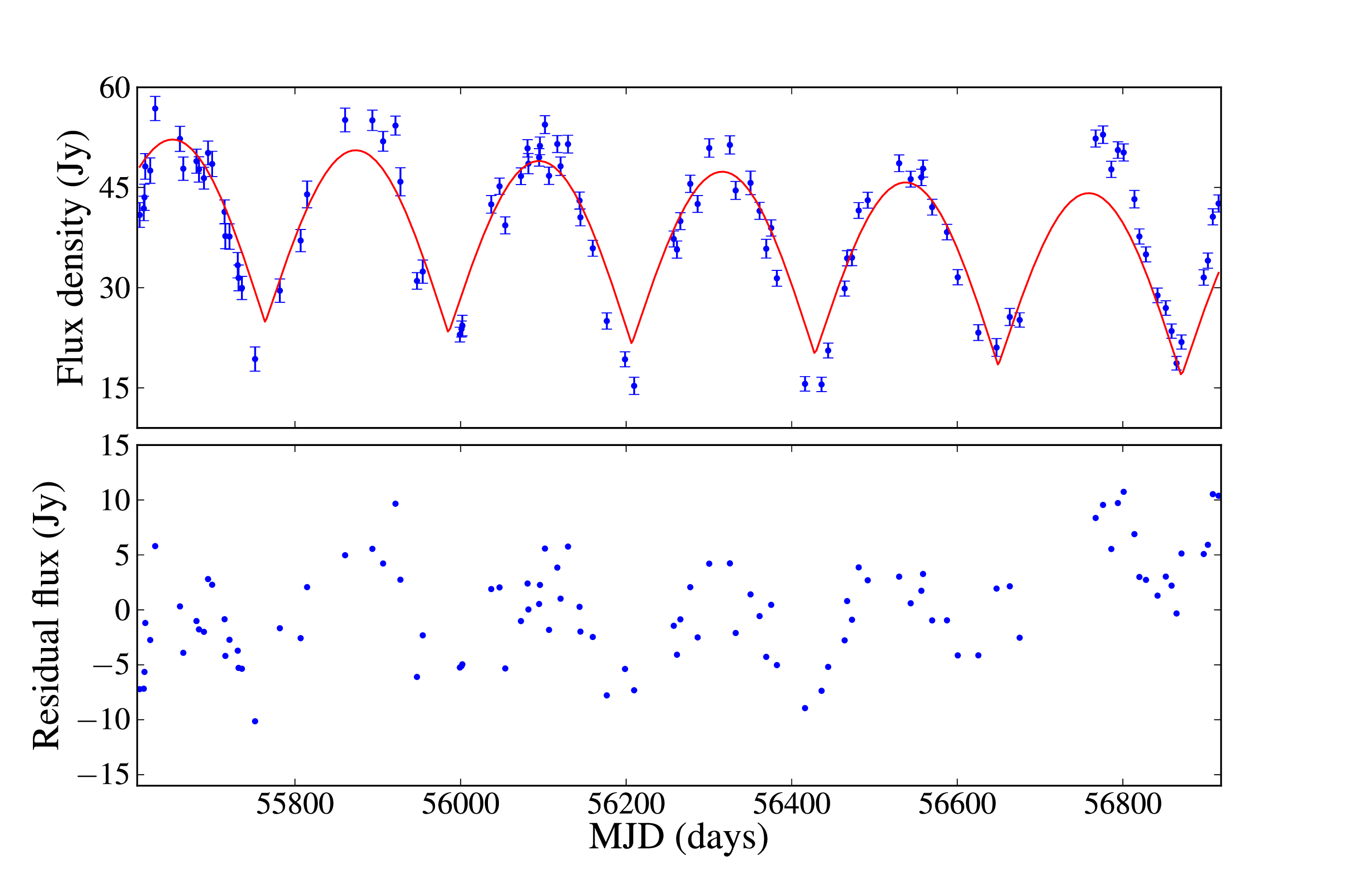}}
\caption{Fitted absolute cosine with 220.0 $\pm$ 0.2 days period to the 1.129 km.s$^{-1}$ time series. The bottom panel is for the residual flux densities after the best fit of an absolute cosine had been subtracted from the 1.129 km.s$^{-1}$ time series.}
\label{fig:G358.460-0.391_abs_cosine}
\end{figure}

\begin{figure}
\centering
\resizebox{\hsize}{!}{\includegraphics[clip]{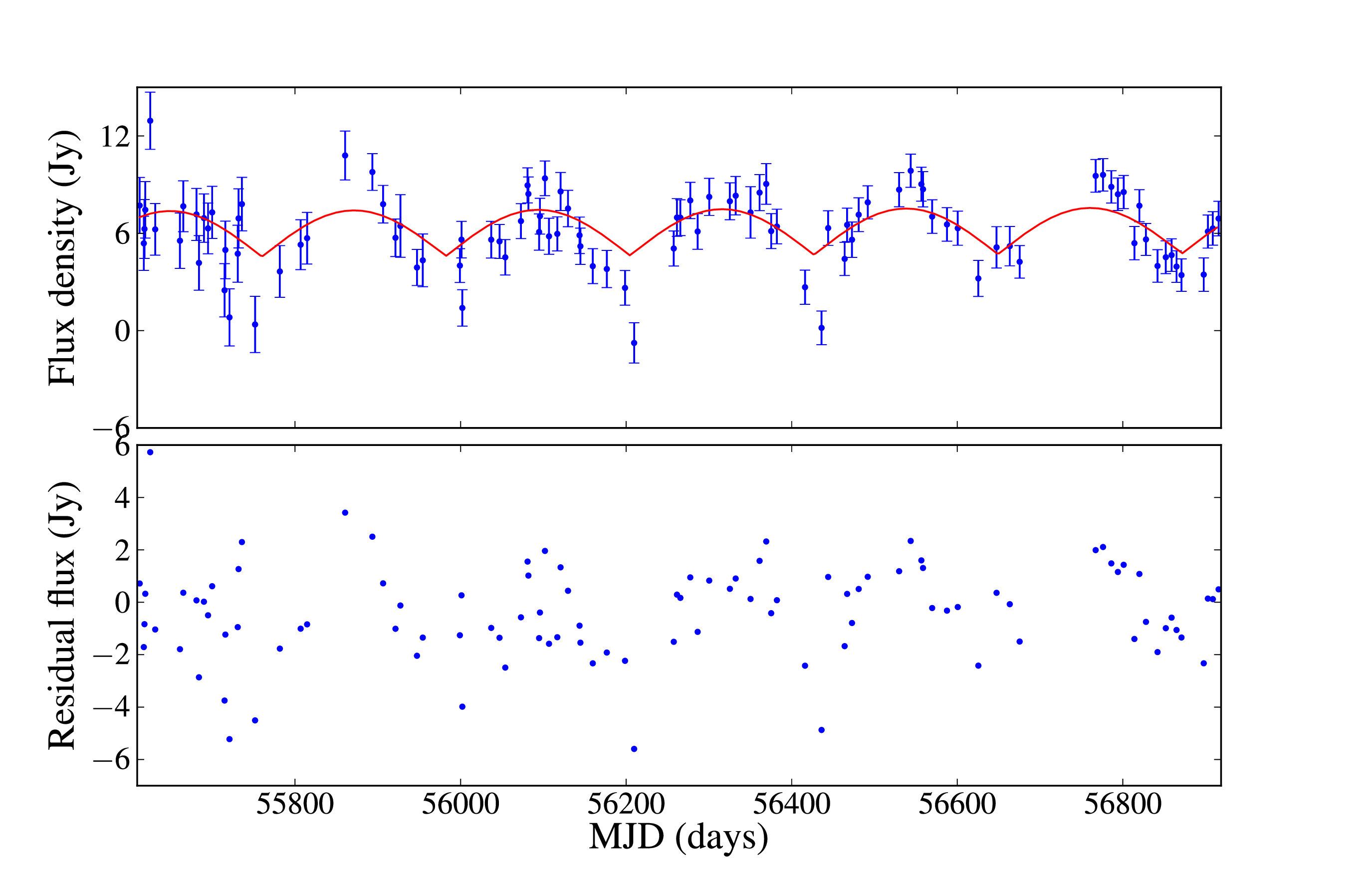}}
\caption{Fitted absolute cosine with 223 $\pm$ 2 day period to the 1.700 km.s$^{-1}$ time series.}
\label{fig:G358.460-0.391_abs_cosine_2nd}
\end{figure}

Visual inspection of the light curves for the masers in G358.460-0.391 suggests that it can be modelled by an absolute cosine function. Figures \ref{fig:G358.460-0.391_abs_cosine} and \ref{fig:G358.460-0.391_abs_cosine_2nd} show such a function fitted to the time series, using the form
\begin{equation}
f(t) = A\left| \cos{\left( \frac{\omega t}{2} + \phi \right) }\right| + mt + c,
\label{eq:absolute_cosine_function}
\end{equation}
where $A$, $\omega$, $t$, $m$ and $c$ are the amplitude, frequency, time, start-phase, gradient and y-intercept, respectively. The fitted parameters were $A$, $\omega$, $\phi$, $m$ and $c$. The G358.460-0.391 waveform is similar to that of G338.93-0.06 \citep{goedhart2004,goedhart2013}. 

\begin{table}
  \caption{The determined periods, gradient (m), mean absolute errors (MAE), mean bias errors (MBE)  chi-square (${\chi}^2$) values and reduced chi-square ($\chi_{red}^2$) values for the optimum fit to the time series.}
  \label{tab:fitted_parameters}
  \begin{center}
  \begin{tabular}{@{}llllll@{}}
  \hline
    Velocity       &  Period & Gradient            & MAE & MBE \\
    (km.s$^{-1}$)  & (days)  &  (Jy.${day}^{-1}$)  &    &     \\
    \hline
1.129              & 220.0 $\pm$ 0.2 &  -0.0066 $\pm$ 0.0004  &  3.5  & -0.2  \\
1.700              & 223   $\pm$ 2   &   0.0002 $\pm$ 0.0003  &  1.4  & 0.4 \\
    \hline

\end{tabular}
  \end{center}
\end{table}

Using this model, the period and its uncertainty can be determined by using chi-square $\chi^2(t,p)$ minimisation. The parameters of the model which best fit the data were determined using Levenberg-Marquardt method \citep{levenberg1944,marquardt1963}. The errors on the best fit parameters were found from the diagonal of the square root of the inverse covariance matrix \citep{press1992}.

The fitted absolute cosine model for the 1.129 and 1.700 km.s$^{-1}$ time series are shown in Figures \ref{fig:G358.460-0.391_abs_cosine} and \ref{fig:G358.460-0.391_abs_cosine_2nd}, respectively. The mean absolute error, MAE, and mean bias error, MBE, \citep{willmott2005} were determined to assess the goodness-of-fit using the residual. The values of MAE and MBE confirm that an absolute cosine function is a good model for the 1.129 and 1.700 km.s$^{-1}$ time series since they are close to zero (Table \ref{tab:fitted_parameters}), although the 1.129 km.s$^{-1}$ time series had a large value owing to a variable peak intensity. 

The gradients, $m$, in Table \ref{tab:fitted_parameters} suggest that there is insignificant to no long term variability over our monitoring window. The implication could be that the origin of periodicity in G358.460-0.391 is stable, unlike G338.93-0.06 which had flux variations at minimum and maximum.

The absolute cosine fit to the G358.460-0.391 time series should not be interpreted as suggesting a specific underlying physical model but it is more to guide the eye. The fit was important in determining the period and its uncertainty, and to test for the significance of the long term linear variability.  

%
\subsection{Maser spots geometry}

\begin{figure}
\centering
\resizebox{\hsize}{!}{\includegraphics[clip]{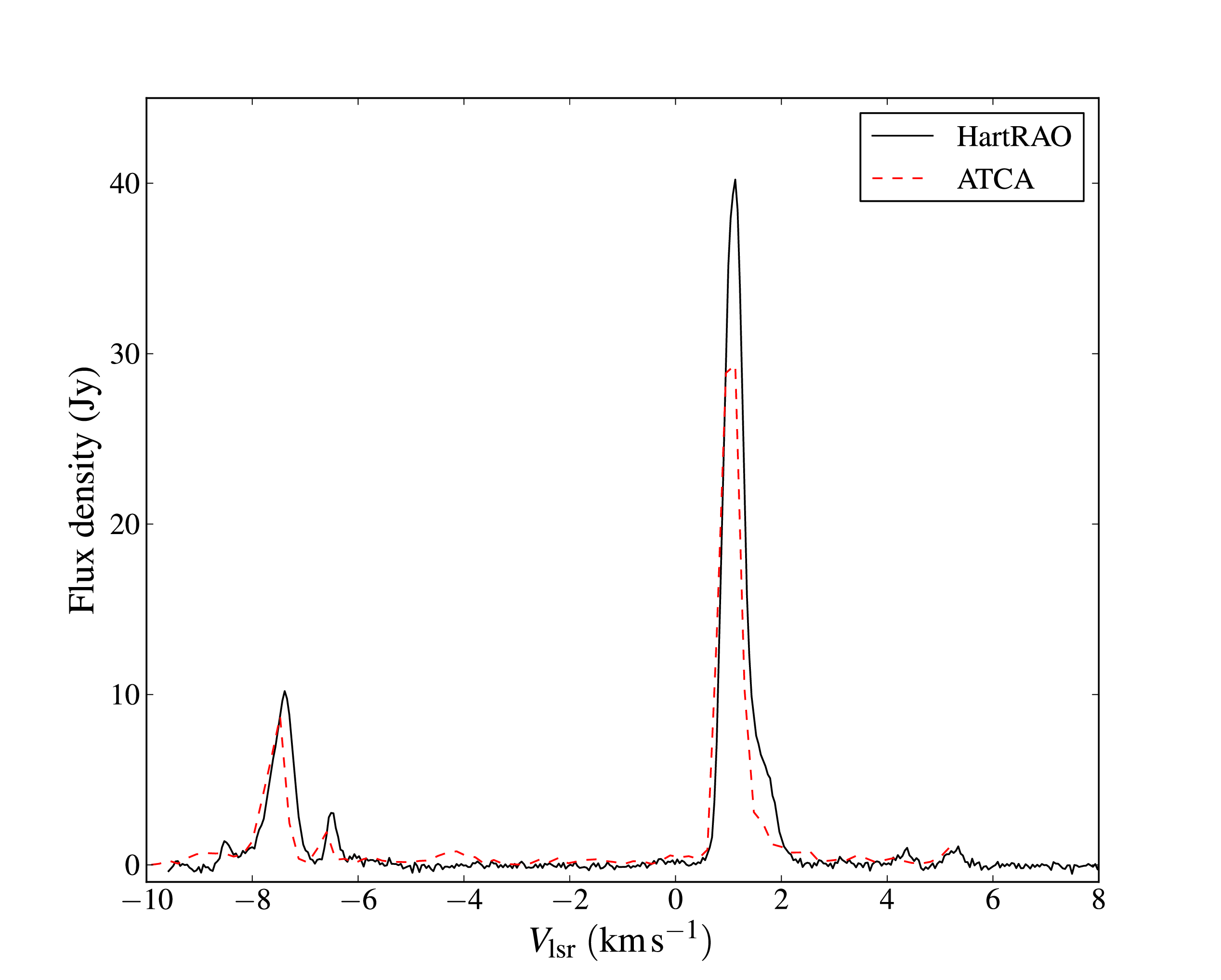}}
\caption{Comparison between average HartRAO spectra and  ATCA Stokes I polarisation spectrum averaged with time and baselines for methanol masers associated with G358.460-0.391 and G358.460-0.393 at 6.7 GHz. The maser features at positive velocities are from G358.460-0.391 and the negative features are for G358.460-0.393.}
\label{fig:G358.460-0.391_atca_spectrum}
\end{figure}

\begin{figure}
\resizebox{\hsize}{!}{\includegraphics[clip]{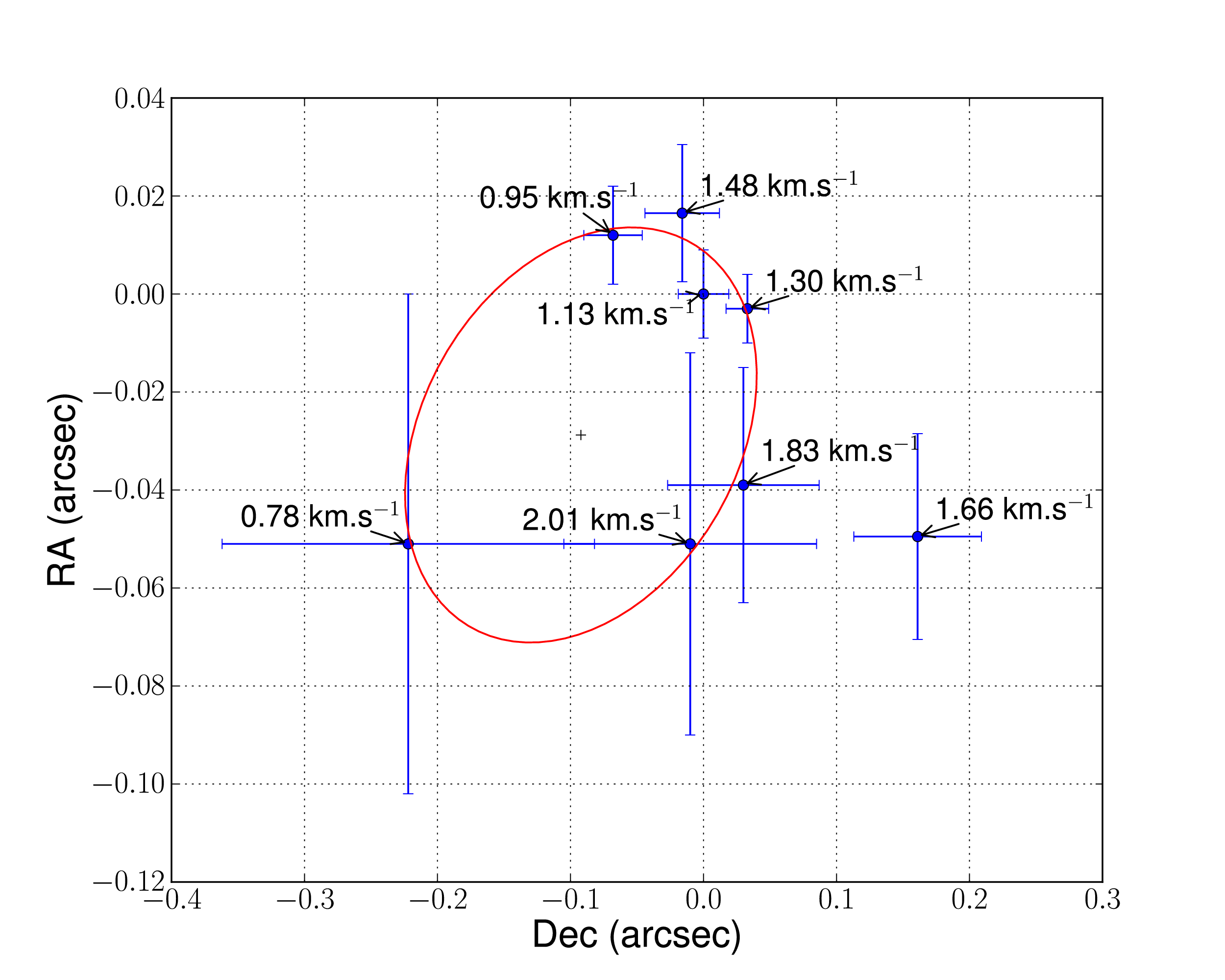}}
\caption{Maser spots map for methanol associated with G358.460-0.391 at 6.7 GHz. The red solid line is the fitted ellipse to the maser spots.}
\label{fig:G358.460-0.391_maser_spots}
\end{figure}

The interferometric observations with ATCA in G358.460-0.391 and G358.460-0.393 produced the spectrum shown in Figure \ref{fig:G358.460-0.391_atca_spectrum}. The spectrum is similar to the one observed by the 26m HartRAO telescope except for the maser features between $4$ and $6$ km.s$^{-1}$. 

The distribution of maser spots in G358.460-0.391 was determined using the five snapshot scans from ATCA and measuring the position of masers in each channel with task SAD in AIPS. The results are shown in Figure \ref{fig:G358.460-0.391_maser_spots} and the spectrum from interferometric observation is shown in Figure \ref{fig:G358.460-0.391_atca_spectrum}. The masers distribution in this region can be fitted with elliptical function. The best fitted semi-major axis, semi-minor axis and eccentricity were 0.13 arcseconds, 0.04 arcseconds and 0.95 respectively. The major diameter of an ellipse is about 0.26 arcseconds.

%
\subsection{Probability of detecting periodic masers}

The sizes of the samples that were monitored have only been given by \citep{goedhart2004} (fifty four sources) and this work (eight sources) which together produced eight detections. These statistics enable the probability of finding periodic methanol masers to be estimated as 0.13. The uncertainty in this estimate was derived both by Monte Carlo simulation and sample size formula for an infinite population \citep{weiss2012}. 

In the Monte Carlo simulation, 5000 methanol maser population sample of which 13 per cent were periodic sources and were randomly distributed in the cells. The probability of detection was calculated from sixty-two randomly selected cells with maser sources. After 10 000 iterations, the standard deviation of the probability of detection from the histogram was used as error estimate. 

The uncertainty in the probability of detection was also derived from the formula for the standard deviation of the probability of detection $\sigma_{\hat{p}}$ for an infinite population, which is given by
\[
\sigma_{\hat{p}} = Z_{\alpha} \sqrt{ \frac{\hat{p} \left( 1 - \hat{p}
\right)}{n}}.
\]
Where $Z_{\alpha}$, $\hat{p}$ and $n$ are the $Z_{\alpha}$ values that has area $\alpha$ to its right under the normal distribution curve, probability of detection and sample size, respectively \citep{weiss2012}. For $\hat{p}=0.13$, $Z_{\alpha}=1$ and $n=62$, then $\sigma_{\hat{p}}=0.04$, which agrees with the uncertainty from the simulation.

However, it may be questioned how well the samples monitored by HartRAO represent the full set of methanol masers in the Milky Way. Both HartRAO samples used the brightness of the methanol masers as a selection criterion. The apparent brightness of a maser is a strong function of distance to the source, and the location of the solar system in the Milky Way could influence the characteristics of the sample. In addition, it has been proposed that brighter methanol masers are associated with later evolutionary phases of massive star formation \citep{breen2010,breen2011}. With these caveats in mind, the estimate of the probability of a maser being periodic as 0.13 $\pm$ 0.04 is a useful working value until more samples are assessed.

%
\section{Discussion}

As already noted earlier, the light curves for G338.93-0.06 and G358.460-0.391 are similar which makes sense to discuss both sources here. Comparison of the light curves of the periodic masers in these two sources with that of the periodic masers in G9.62+0.20E suggests that the underlying mechanism in G338.93-0.06 and G358.460-0.391 must be quite different from that in G9.62+0.20E. In the latter case, the periodic masers has the nature of a flaring behaviour starting from a low ``equilibrium state'', whereas in the case of G338.93-0.06 and G358.460-0.391 the ``equilibrium state'' seems to be the high state of maser emission. In fact, the light curves of G338.93-0.06 and G358.460-0.391 very much resemble that of some eclipsing binary systems \citep[e.g.][]{mcvean1997,kaluzny2006,clausen2007}. 

\begin{figure}
\resizebox{\hsize}{!}{\includegraphics[clip]{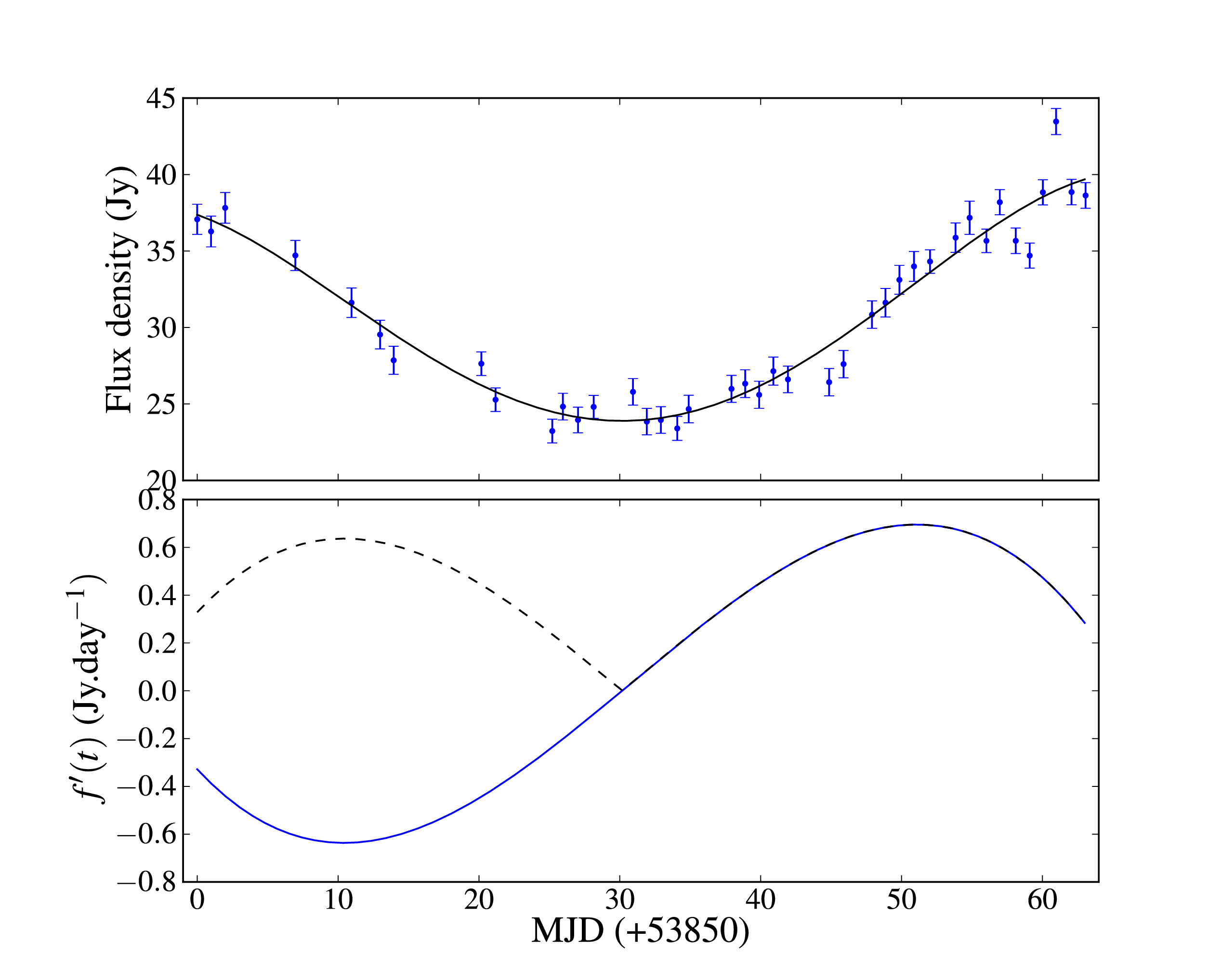}}
\caption{The top panel is the time series for the 6.7 GHz methanol masers associated with G338.93-0.06 between 53904-53967 MJD fitted with a fifth order polynomial (solid line). The bottom is derivative, $f^{\prime}(t)$, (dash line) and absolute of the derivative, $\left| f^{\prime}(t) \right|$, (solid) of the fifth order polynomial fitted to the time series in the top panel.}
\label{fig:G338.93-0.06_around_minimum}
\end{figure}

One of the properties of the light curves of eclipsing binary systems is that the decreasing and increasing sides adjacent to a minimum is rather symmetric. For G338.93-0.06 and G358.460-0.391 there is basically only one instance where the decrease towards a minimum and the subsequent increase  had been sampled rather well and this is for the minimum close MJD 53900 for G338.93-0.06 \citep[e.g. see][]{goedhart2007}. To determine to what degree there is symmetry around this minimum, a fourth order polynomial was fitted to the data from 30 days before the minimum to 30 days after the minimum. The fit is shown in the upper panel of Figure \ref{fig:G338.93-0.06_around_minimum}. In the bottom panel is the derivative of the fourth order polynomial and their absolute values are shown. Inspection of both panels suggests that the minimum occurred around MJD 53935. Inspection of the derivative further suggests remarkable symmetric behaviour of the light curve on both sides of the minimum. Although this symmetric behaviour is not a proof for an eclipsing mechanism, it certainly points to a mechanism that results in the symmetric behaviour of the masers around the minima. Figure \ref{fig:G338.93-0.06_around_minimum} also shows that G338.93-0.06 spend few days at the minimum ($\sim$ 5 days), but this does not automatically translate into G358.460-0.391 not having a sharp minimum because they are different sources. This can be resolved by monitoring the source at least once every two days. 

Assuming for the moment that G338.93-0.06 and G358.460-0.391 are binary systems, given the periods derived for the two sources (133 and 220 days) and, assuming a total stellar mass of 20$M_\odot$, the semi-major axes for the two systems will be 1.38 AU and 1.94 AU respectively. If the behaviour of the periodic masers in these two sources is indeed due to some kind of eclipsing event, it is clear that it cannot be the maser itself that is eclipsed by the secondary star since the masing regions are located significantly further from the massive star. A possible alternative scenario is that the secondary star is eclipsing some of the ionizing radiation from the primary (high mass) star that propagates in the direction of that part of the HII region against which the maser is projected and thus affects the ionization and therefore the flux of free-free seed photons. In this regard, note that the secondary star might well be a lower mass star. For example, a few solar masses for which the radius is significantly larger than that of the equivalent main-sequence star, e.g., a 3$M_\odot$ to 4$M_\odot$ protostar has a radius of about 10$R_\odot$ \citep{palla1990}. If the case of, e.g. G338.93-0.06 with semi-major axis of 1.38 AU is considered, then for eccentricities of 0.8, 0.85, and 0.9 the periastron distances are respectively 0.28 AU, 0.21 AU, and 0.14 AU. An object with a radius of 10$R_\odot$ will subtend solid angles of 0.09 sr, 0.16 sr, and 0.36 sr respectively for these distances as viewed from the primary star. From a purely geometric point of view the circular ``shadow'' casted by such an object will have a radius $r = \sqrt{\Omega d^2/\pi}$ at a distance $d$ from the primary star. For typical distances of a few 100 AU for the size of the associated hyper-compact HII region, $r$ is of the order of tens of AU which is significantly larger than the projected size of the maser spot against the HII region. The shadowing effect will obviously not be perfect since diffuse ionizing photons in the HII region will still enter the shadowed region.

The above scenario might qualitatively explain some of the properties of the observed light curves of G338.93-0.06 and G358.460-0.391. However, the light curves of these two sources clearly suggest that whatever the underlying mechanism, it seems to affect the masers most of the time since there is not a clear stable state. It might be very difficult to explain this aspect within the framework of the above described eclipsing scenario. It is also not clear how the light curves of these two sources can be explained within the framework of the periodicity being due to changes in the masing region itself, for example, if it is due to changes in the dust temperature as proposed by some authors. 

It seems that there are at least two categories of periodic maser sources on the basis of the shape of the light curves. The first group consists of G9.62+0.20E,  G22.357+0.066 and G37.55+0.20 for which the flare profile seems to be rather well described in terms of the increase in the level of ionization followed by the recombination of the plasma. The second group consists of G338.93-0.06 and G358.460-0.391 with light curves similar to that of eclipsing binary system. Inspection of the light curves of other periodic masers {\it suggests} that G339.62-0.12 might also belong to the second group while the light curve of G331.13-0.24 shows some similarities to that of G9.62+0.20E. It surely is an interesting question whether all light curves can be grouped into these two categories or whether other categories can also be identified.

Together with the previous discoveries of the periodic masers, there are now eleven such sources. This raises the question of how many more there may be and therefore, if any further searches are warranted. In all the previous surveys conducted at HartRAO, the main criterion for selecting candidate sources was that the flux density of the brightest maser feature in a specific source must be greater than ten Jansky in order to have sufficient signal-to-noise ratio. Using the detected sources of the MMB survey catalogues I, II, III and IV \citep{caswell2010,green2010,caswell2011,green2012}, 262 sources were found to meet the criteria set above. Applying the probability to find a periodic maser, it implies that there may be 34 $\pm$ 10 periodic sources in total. Given that eight masers had already been  discovered, it means that a further twenty-six or so periodic masers might be discovered among the remaining 200 brighter methanol masers.

\section{Summary}

Eight of the brightest class II methanol masers from the MMB survey catalogues I and II were monitored with the 26m HartRAO radio telescope. Of the eight sources, only the 6.7 GHz maser of G358.460-0.391 shows periodic variations. The periods determined by the Lomb-Scargle, epoch-folding, Jurkevich and absolute cosine fitting techniques agree, but it is the latter method that produced the best estimate of the period and its uncertainties. The determined period for G358.460-0.391 was 220.0 $\pm$ 0.2 days. Accurate estimation of the period and its error are essential for testing the stability of the period and testing for the potential evolution of the period and the level of intrinsic variation from cycle to cycle. 

It is important to search for more periodic class II methanol masers in order to understand the origin of the periodicity, the types of waveforms and the frequency with which each occurs, to redefine or understand the period range in association with their environment, to understand masers and their surroundings. A careful analysis of the waveforms offers a possible method of distinguishing between scenarios. 

From the two surveys, it was estimated that the probability of finding a periodic methanol maser in a given sample is 0.13 $\pm$ 0.04, for the given source selection criteria described above. This leads to a proposal of the expected 34 $\pm$ 10 periodic masers in the MMB survey catalogues I, II, III and IV. From these catalogues, about twenty-six periodic methanol masers are still expected to be discovered, which suggests that it might be worth searching for more in the MMB survey catalogues. 

Maser numerical modelling is also a vital tool in understanding the factors that could be causing periodic variability. It could help answer questions such as whether only changes in the infrared pump photon flux can cause these variations or if the seed photons at the microwave line frequency are also involved.

\section*{Acknowledgments}
Jabulani Maswanganye would like to express his gratitude for the bursaries from the National Astrophysics Space Science Programme, Hartebeesthoek Radio Astronomy Observatory, the North-West University doctoral scholarships and the South African SKA Project via the NRF. The authors would like to dedicate this article to Dr. M.J. Gaylard, who never saw the end of this final version.

\bsp

\label{lastpage}


\begin{thebibliography}{99} 
\bibitem[\protect\citeauthoryear{Andrae et al.}{2010}]{andrae2010} Andrae R., Schulze-Hartung T., Melchior P., 2010, ApJ, preprint (arXiv:1012.3754)
\bibitem[\protect\citeauthoryear{Araya et al.}{2010}]{araya2010} Araya E.D. et al., 2010, ApJ, 717, L133
\bibitem[\protect\citeauthoryear{Breen et al.}{2010}]{breen2010}Breen S.L., Ellingsen S.P., Caswell J.L., Lewis B.J., 2010, MNRAS, 401, 221
\bibitem[\protect\citeauthoryear{Breen et al.}{2011}]{breen2011} Breen S.L. et al., 2011, ApJ, 733, 80
\bibitem[\protect\citeauthoryear{Caswell et al.}{2010}]{caswell2010} Caswell J.L. et al., 2010, MNRAS, 404, 1029
\bibitem[\protect\citeauthoryear{Caswell et al.}{2011}]{caswell2011} Caswell J.L. et al., 2011, MNRAS, 417, 1964
\bibitem[\protect\citeauthoryear{Clausen at al.}{2007}]{clausen2007} Clausen J.V. et al., 2007, A\&A, 461, 1065
\bibitem[\protect\citeauthoryear{Davies}{1990}]{davies1990} Davies S. R., 1990, MNRAS, 244, 93
\bibitem[\protect\citeauthoryear{Fujiswa et al.}{2014}]{fujiswa2014} Fujisawa K. et al., 2014, preprint (arXiv:1405.5972v1) 
\bibitem[\protect\citeauthoryear{Green et al.}{2009}]{green2009} Green J.A. et al., 2009, MNRAS, 392, 783
\bibitem[\protect\citeauthoryear{Green et al.}{2010}]{green2010} Green J.A. et al., 2010, MNRAS, 409, 913
\bibitem[\protect\citeauthoryear{Green et al.}{2012}]{green2012} Green J.A. et al., 2012, MNRAS, 420, 3108
\bibitem[\protect\citeauthoryear{Goedhart et al.}{2003}]{goedhart2003} Goedhart S., Gaylard M.J., van der Walt D.J., 2003, MNRAS, 339, L33
\bibitem[\protect\citeauthoryear{Goedhart et al.}{2004}]{goedhart2004} Goedhart S., Gaylard M.J., van der Walt D.J., 2004, MNRAS, 355, 553
\bibitem[\protect\citeauthoryear{Goedhart et al.}{2007}]{goedhart2007} Goedhart S., Gaylard M.J., van der Walt D.J., 2007, IAU Symposium, 242, 97
\bibitem[\protect\citeauthoryear{Goedhart et al.}{2009}]{goedhart2009} Goedhart S., Langa M.C., Gaylard M.J., van der Walt D.J., 2009, MNRAS, 398, 995
\bibitem[\protect\citeauthoryear{Goedhart et al.}{2013}]{goedhart2013} Goedhart S., Maswanganye J.P., Gaylard M.J., van der Walt D.J., 2013, MNRAS, 437, 1808
\bibitem[\protect\citeauthoryear{Inayoshi et al.}{2013}]{inayoshi2013} Inayoshi K., Sugiyama K., Hosokawa T., Motogi K., Tanaka K.E.I., 2013, ApJ, 769, L20
\bibitem[\protect\citeauthoryear{Jurkevich}{1971}]{jurkevich1971} Jurkevich I., 1971, Ap\&SS, 13, 154
\bibitem[\protect\citeauthoryear{Kaluzny et al.}{2006}]{kaluzny2006} Kaluzny J., Thompson I.B., Krzeminski W., Schwarzenberg-Czerny A., 2006, MNRAS, 365, 548
\bibitem[\protect\citeauthoryear{Kidger et al.}{1992}]{kidger1992} Kidger M.R., Takalo L., Sillanp\"{a}\"{a} A., 1992, A\&A, 264, 32
\bibitem[\protect\citeauthoryear{Leahy et al.}{1983}]{leahy1983} Leahy D.A. et al., 1983, ApJ, 266, 160
\bibitem[\protect\citeauthoryear{Levenberg}{1944}]{levenberg1944} Levenberg K., 1944, The Quarterly of Applied Mathematics, 2, 164
\bibitem[\protect\citeauthoryear{Lomb}{1979}]{lomb1976} Lomb N.R., 1976, ApSS, 39, 447
\bibitem[\protect\citeauthoryear{Marquardt}{1963}]{marquardt1963} Marquardt D.W., 1963, J. Soc. Ind. Appl. Math., 11, 431
\bibitem[\protect\citeauthoryear{McVean et al.}{1997}]{mcvean1997} McVean J.R., Milone E.F., Mateo M., Yan L., 1997, ApJ, 481, 782
\bibitem[\protect\citeauthoryear{Ott et al.}{1994}]{ott1994} Ott M. et al., 1994, A\&A, 284, 331
\bibitem[\protect\citeauthoryear{Palla \& Stahler}{1990}]{palla1990} Palla F., Stahler S.W., 1990, ApJ, 360, L47
\bibitem[\protect\citeauthoryear{Parfenov \& Sobolev}{2014}]{parfenov2014} Parfenov S. Yu, Sobolev A.M., 2014, MNRAS, preprint (arXiv:1407.7708v1) 
\bibitem[\protect\citeauthoryear{Press et al.}{1989}]{press1989} Press W.H., Rybicki G.B., 1989, ApJ, 338, 277
\bibitem[\protect\citeauthoryear{Press et al.}{1992}]{press1992} Press W.H., Teukosky S.A., Vetterling W.T., Flannery B.P., 1992, Numerical Recipes The Art of Scientific Computing, Cambridge University Press, 3rd Edition, pp. 1003
\bibitem[\protect\citeauthoryear{Scargle}{1982}]{scargle1982} Scargle J.D., 1982, ApJ, 263, 835
\bibitem[\protect\citeauthoryear{Sobolev et al.}{2007}]{sobolev2007} Sobolev A.M. et al., 2007, IAUS, 242, 81
\bibitem[\protect\citeauthoryear{Stellingwerf}{1978}]{stellingwerf1978} Stellingwerf R.F., 1978, ApJ, 224, 953
\bibitem[\protect\citeauthoryear{Szymczak et al.}{2011}]{szymczak2011} Szymczak M., Wolak P., Bartkiewicz A., van Langevelde H.J., 2011, A\&A, 531, L3
\bibitem[\protect\citeauthoryear{van der Walt}{2005}]{vanderwalt2005} van der Walt D.J., 2005, MNRAS, 360, 153
\bibitem[\protect\citeauthoryear{van der Walt et al.}{2009}]{vanderwalt2009} van der Walt D.J., Goedhart S., Gaylard M.J., 2009, MNRAS, 398, 961
\bibitem[\protect\citeauthoryear{van der Walt}{2011}]{vanderwalt2011} van der Walt D.J., 2011, ApJ, 141, 152
\bibitem[\protect\citeauthoryear{Weiss}{2012}]{weiss2012} Weiss N.A., 2012, Elementary statistics, 8th ed, Addison--Wesley, Boston, p. 445 
\bibitem[\protect\citeauthoryear{Willmott \& Matsuura}{2005}]{willmott2005} Willmott C.J., Matsuura K., 2005, Climate Res, 30, 79
\bibitem[\protect\citeauthoryear{Xie et al.}{2008}]{xei2008} Xie G.Z., Yi T.F., Li H.Z., Zhou S.B., Chen L.E., 2008, ApJ, 135, 2212 


\end{thebibliography}
\end{document}